\documentclass[sigconf]{acmart}

\usepackage{nameref,hyperref}
\usepackage{comment}
\usepackage{color}
\usepackage{hyphenat}
\usepackage{xcolor}
\usepackage{multirow}
\usepackage{longtable}
\usepackage{array}
\usepackage{breakcites}
\usepackage{supertabular}
\usepackage{todonotes}
\usepackage{hyperref}
\usepackage{tikz}
\usetikzlibrary{mindmap}
\usetikzlibrary{positioning}

\newcolumntype{R}[1]{>{\raggedleft\arraybackslash }p{#1}}
\newcolumntype{L}[1]{>{\raggedright\arraybackslash }p{#1}}
\newcolumntype{C}[1]{>{\centering\arraybackslash }p{#1}}

\newcommand{\specialcell}[2][c]{%
    \begin{tabular}[#1]{@{}c@{}}#2\end{tabular}}

\AtBeginDocument{%
  }

\setcopyright{none}
\acmDOI{}
\acmISBN{}  
\acmYear{}
\acmConference[LIMITS '24]{Tenth Workshop on Computing within Limits}{June 18--19, 2024}{Online}

\begin{document}

\title[How digital will the future be? ]{ How digital will the future be? \\Analysis of prospective scenarios}
\author{Aurélie Bugeau}
\email{aurelie.bugeau@u-bordeaux.fr}
\affiliation{\institution{Univ. Bordeaux, CNRS, Bordeaux INP, LaBRI, UMR5800, IUF}   \city{F-33400 Talence} \country{France}
}

\author{Anne-Laure Ligozat}
\email{anne-laure.ligozat@lisn.upsaclay.fr}
\affiliation{\institution{Université Paris-Saclay, CNRS, ENSIIE, Laboratoire Interdisciplinaire des Sciences du Numérique}   \city{ 91400 Orsay} \country{France}}

\renewcommand{\shortauthors}{A. Bugeau, A.-L. Ligozat}

\begin{abstract}
Within the context of climate change, many prospective studies generally encompassing all areas of society, imagine possible futures to expand the range of options. The role of digital technologies within these possible futures is rarely specifically targeted. Which digital technologies and methodologies do these studies envision in a world that has mitigated and adapted to climate change?  In this paper, we propose a typology for scenarios to survey digital technologies and their applications in 14 prospective studies and their corresponding 35 future scenarios. Our finding is that all the scenarios consider digital technology to be present in the future. We observe that only a few of them question our relationship with digital technology and all aspects related to its materiality, and none of the general studies envision breakthroughs concerning technologies used today. Our result demonstrates the lack of a systemic view of information and communication technologies. We therefore argue for new prospective studies to envision the future of ICT. 
\end{abstract}

\keywords{future of ICT, prospective studies, future scenarios, sustainable ICT}

\maketitle

\section{Introduction}

\begin{figure}
    \centering
    \includegraphics[width=.3\textwidth]{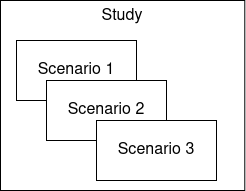}
    \caption{Each study may include several scenarios, representing alternative futures}
    \label{fig:study_includes_scenarios}
\end{figure}

In the context of climate change, many prospective studies have been proposed in recent years to envision possible futures via both narrative and quantitative formats.  Different techniques may be used in future study: reflection, simulation and modeling, monitoring or behavioral~\cite{Mankoff2013}. Prospective studies often present multiple scenarios (see Figure~\ref{fig:study_includes_scenarios}) resulting from different societal or technical hypotheses, each shaped by many variables, such as economic growth, well-being, sobriety, low-carbon energy, population, or uncertain events. They help cope with uncertainties and challenges by expanding the horizons of futures that can be imagined~\footnote{https://www.geoffmulgan.com/post/social-imagination}. Future studies can help being better prepared for meeting future challenges, enabling the invention of a better future, or increasing the long-term impact of our work~\cite{Mankoff2013}. They are actively used by policy designers. 
The prospective scenarios are neither predictions nor forecasts but rather overviews of possible future paths~\cite{Arup}. Studies often include a baseline business-as-usual scenario used as a reference.

The LIMITS community has a long history of working on future studies and on ICT futures~\cite{Mankoff2013,Penzenstadler2014,Nardi2016,Tanenbaum2016,  eriksson2018meeting, pargman2019future}. Among the many research topics they address, the community questions what will research look like in the context of scarcity, how design fiction can challenge business-as-usual and pluralize the future, how fiction allows to adopt a range of different intellectual commitments and values about the future, and how to ascend from future studies to the concrete. 
Authors of \cite{nardi2018computing} underline the role that computing has to play in responding to global limits and in shaping a society that meaningfully adapts to them. They nevertheless state that many computing researchers and practitioners in practice assume there is only one possible likely future, one very much like
the present. 

As computer scientists, we wanted to question the place of digital technologies in studies that have recently been designed to support policy and guide action towards sustainability, and to develop the imagination in this context. Through a preliminary analysis of several scenarios, we observed that they always consider Information and Communication Technologies (ICT) to be present in the future. However, the evolution of the ICT sector and its role in other sectors vary between scenarios. In this paper, we propose to dig further into a set of prospective studies to summarize the future of ICT they envision. More specifically, the objective of this work is to answer the following questions: 
\begin{itemize}
\item What ICT is present in the scenarios, i.e., which infrastructure, amount of data, and technologies?
\item To what extent does ICT vary among scenarios?
\item What applications relying on ICT are described?
\end{itemize}

To answer these questions, we analyzed 14 studies and their corresponding 35 scenarios, and we focused on the evolution of digital technologies and their areas of applications.  We do not pretend to make an exhaustive review of existing scenarios. However, we tried to cover several scenario types within a spectrum of narrative and quantitative formats and present positive or negative visions of the future. We identified a limitation of our study, namely the selection of OECD countries-centered studies, which we discuss in Section~\ref{sec:discussion}. 

Our contributions are to:
\begin{itemize}
    \item Analyze digital technologies within scenarios according to a proposed set of variables organized in a comprehensive typology.
    \item Highlight common points and differences between scenarios.
    \item Identify challenges that should be addressed as research questions to enable or avoid these scenarios.
    \item Identify some missing aspects regarding ICT in prospective scenarios. 
\end{itemize}
Section~\ref{sec:related} presents existing work on ICT concerning future scenarios. Section \ref{sec:description} explains how the studies were selected and discusses the variables used to analyze the scenarios. 
Section \ref{sec:IT} details the results for ICT variables. A focus is made on the digital technologies and area of applications explicitely mentioned in the scenarios. Lastly, in section \ref{sec:discussion}, we discuss the omitted features in the studies that we identified.

\section{Related works}~\label{sec:related}

We found only very few studies that compare and analyze the place of digital technologies in prospective scenarios. In this section, we describe the works that we found related to ours. 

Several authors estimate future ICT energy consumption and carbon footprint \cite{andrae2019prediction,su10093027}. The projections depend on two main variables: ICT equipment and services efficiency, and demand for ICT (via possible rebound effects) \cite{freitag2021}. Our goal here is not to quantify the future consumption of ICT but rather to analyze the possible places of ICT in the future and to explicitly state the implications it could have on first-, second-, and third-order effects~\cite{hilty2006relevance} of the sector.

In~\cite{Erdmann2010}, the authors propose a scenario analysis of the future impacts of ICT applications on GHG emissions. ICT being a driver of new lifestyles, structural changes, and economic growth, studying its effects is complex. The authors start by reviewing existing macroeconomics on ICT and GHG emissions and conclude that, apart from a few studies, previous works show unambiguous overall reductions of GHG emissions due to ICT. The second part focuses on building specific macroeconomic ICT impact assessment models by looking at ICT's first-, second-, and third-order effects on the environment. It extends a previous study~\cite{erdmann2004future} by focusing on data uncertainty, requiring further research, and future uncertainty indicating a need for political action. 
The proposed macroeconomic ICT impact assessment model  is applied to the three scenarios from~\cite{erdmann2004future}. These scenarios designed in 2003 describe alternative future
courses of ICT until 2020. The scenario analysis from ~\cite{Erdmann2010} concludes with the necessity of incorporating ICT in a general macroeconomic model to quantify rebound effects of energy efficiency measures and the need for a large geographical scope for a complete impact assessment of future scenarios.

Also focusing on assessing future scenarios, authors in \cite{faure2017methods} review methods for assessing future scenarios' environmental and social sustainability impacts. They consider three types of scenarios according to the classification of \cite{borjeson2006scenario}, i.e., predictive, explorative, and normative scenarios. A particular focus is put on sustainability assessments of scenarios in the ICT sector, an example of a rapidly developing sector. The study concludes that the topic of assessment requires further attention if scenarios are to be used to support policy and guide action towards sustainability.  In this paper, we do not question the environmental and societal impacts assessment of the scenarios but rather compare scenarios based on the imaginaries and research  questions they may drive.


Questioning the evolution in the vision for ICT in European policies, \cite{Andok2022} proposes an analysis of digital futures in different European Union reports. The author reviews three EU reports on digital futures. The first one, \cite{EU2014}, published in 2014, is very techno-optimistic in every aspect of the society. The second (post-COVID pandemic) report~\cite{EU2021} from 2021 is slightly less optimistic and mentions vulnerabilities and disinformation. It also discusses sustainability, mentioning how digital technologies can contribute to EU Green Deal objectives. The third report~\cite{EU2022} gives short-term objectives regarding cyber-security after the beginning of the Russian-Ukrainian war. From this analysis, Andok highlights the shift from opportunities only to dangers and threats, and from 40 years perspective visions to shorter ones (10 years or even less). 

Finally, in~\cite{Penzenstadler2014}, the authors present and analyze a compilation of fictional abstracts written by researchers attending the ICT4S conference. They cluster the abstracts according to four major aspects (infrastructure, society, food, and waste), before stating major changes suggested in technological, economic, and social dimensions. This work is related to ours as it aimed at finding links between different scenarios.

\section{Methodology }
\label{sec:description}
This section describes the 14 studies (section~\ref{sec:scenarios}) we chose to review. From these studies and the state of the art on prospective analysis, we identified features that can influence the evolution of ICT within the different scenarios and organized them in a typology as described in  section~\ref{sec:typology}. The 14 studies were then classified according according to the proposed methodology. While we briefly provide first results of the analysis in this section, next section provides a detailed analysis on the representation of ICT the scenarios.

\subsection{Studies selection}\label{sec:scenarios}

In this section, we present how we chose the 14 studies. We give a short overview of the considered studies in Appendix \ref{sec:studies}.

We first selected studies we already knew, then searched for others online. We tried to find studies with various: spatial perimeters (which countries are considered), domains considered (ICT -specific or not), and methodologies (narratives vs quantitative studies). The last criterion we used is the publication year.

\paragraph{Topic} All studies build upon climate change context. We refer to \textit{general studies} as the ones that regard all areas of society. We started our analysis with such general studies~(e.g. \cite{ademe2021transition, IPCC, Shift, Arup}). In this work, nine studies are general. Additionally, we selected two studies focusing on energy in France in 2050~\cite{RTE,France2072}. Finally, obviously, we have included in our work studies that specifically target information and communication technologies:  \cite{creutzig2022,CNIL,deron2022}. Although we did not select the studies according to the presence of ICT in them, all of them actually mention ICT. 

\paragraph{Publication year}  We limited ourselves to recent studies. The oldest one in our deep analysis was published in 2018. We found very few notions of digital technologies in the oldest general scenarios. For example, in ~\cite{OECD}, we found only one mention of data access, smart metering, and a suggestion to make chemical information widely available through the Internet thanks to the development of new and innovative computer models. Another example of a scenario that we purposely did not include in this work is the ICT-specific from \cite{erdmann2004future}. It aims to explore (qualitatively) and assess (quantitatively) how ICT will influence environmental sustainability between 2003 and 2020. 

\paragraph{Spatial scale} As we are located in France, we naturally started our review with the most recent prospective works made in this country: \cite{ademe2021transition,negaWatt,SNBC,RTE,Shift,France2072, CNIL}. We then opened to studies targeting the future of EU: \cite{EUgreendeal,Eionet}.
Finally, we selected studies for the entire world:~\cite{IPCC, Arup, DDC, creutzig2022}. One ICT-specific study focuses on Quebec~\cite{deron2022}. The studies chosen thus mostly target OECD countries, which is a limitation of our work.

\paragraph{Data} With this criterion, we differentiate  qualitative only from more quantitative studies. 
Within qualitative, we include some entirely narrative studies: \cite{Eionet,Arup,DDC}. They include stories of people living in 2050 enabling us to look at what life could more concretely look like, and resemble science fiction. The study \cite{DDC} is in the form of a future scenario toolkit. Many other toolkits\footnote{\url{https://oecd-opsi.org/toolkits/?\_toolkit\_discipline\_or\_practice=futures-and-foresight}} exist but were not studied here as none are ICT-centered, and we considered these toolkits to be redundant with the ones already studied. 
Conversely, some studies are much more quantitative and try to estimate GHG emissions by 2050 (e.g., \cite{RTE} is mainly based on numerical models).

\subsection{Typologies describing the studies and scenarios}
\label{sec:typology}

Each study has its way of designing possible futures, so capturing the diversity of the scenarios is complex~\cite{van2003updated}. Several works have proposed scenario typologies~(e.g.~\cite{ducot1980typology, van2003updated,crawford2019comprehensive}) to capture this diversity. These typologies aim to categorize scenarios by answering the questions of why (project goal), how (process design), and what (scenario content)~\cite{van2003updated}. However, more analysis is needed to understand more deeply why scenarios end up being so similar or different. Scenarios result from hypotheses and parameters evolutions regarding economy, lifestyles and governance, which are not always explicit.

In this work, we focus on the place and role of ICT in future perspectives. We are therefore interested in comparing the different scenarios from our selected studies according to features that influence the place of ICT.

Authors~\cite{erdmann2004future, Erdmann2010} distinguish external variables that include GDP, population, labor demand, electricity prices, and other variables that may influence first-, second-, and third-order impacts of ICT. Examples of variables impacting first-order impacts are technological trends from large to small devices, household internet penetration, or energy efficiency. Variables for second-order impacts include energy efficiency or materials potentials for other sectors of activity. Finally, variables for third-order impacts are elasticity of demand, average personal transport time, or some economic behavior. The authors use all these variables to analyze the three scenarios from~\cite{erdmann2004future}. These variables focus on impacts of ICT but  do not give a clear picture of what ICT is present in the scenarios and the place of ICT if the futures.

Following these works, we propose our own typology and variables to analyze ICT in prospective scenarios. We differentiate three levels of features to characterize every scenario (Table \ref{tab:typology}). \begin{table*}[h!]
\setlength{\tabcolsep}{0.2cm} 
    \centering
   \resizebox{\textwidth}{!}{%
    \begin{tabular}{L{3.05cm}L{3.35cm}C{9.2cm}}
         \hline
         \multirow{4}{*}{Goal and design}& Level: & Study \\
         & Objective:& General overview of the study \\ 
         &  Features overview:& Project goal, content and complexity\\
         & Reference: & \cite{van2003updated} \\
         \hline
        
         \multirow{3}{*}{Societal variables}& Level: & Scenario \\
         & Objective:& non ICT-specific variables that influence scenarios \\ 
         &  Features overview:& Economy, lifestyle, demography, governance\\
         \hline
         
         \multirow{4}{*}{ICT variables}& Level & Scenario \\
         & Objective& ICT-specific variables present in scenarios \\          &  Features overview& Infrastructure, Applications, Technologies \\
         \hline
    \end{tabular}
    }
    \caption{Overview of typology and parametric variables used to compare scenarios}
    \label{tab:typology}
\end{table*}

The first one is the general typology from  \cite{van2003updated}, that we call "goal and design". It enables to view how the study was conducted and its perimeter. The second one is "Societal variables", which concern hypotheses for society pathways regarding lifestyle, economy, governance, demography, and equality.  We propose variables freely adapted from ~\cite{ademe2021transition}. When analyzing the scenarios, we observed that the variables of economic growth, sobriety, the importance of the state, and the collaboration between countries were discussed in all general studies and drive scenarios. These societal variables impact the kind of ICT envisioned; for example, a scenario with solid states may favor centralization of digital infrastructure and services.

Finally, after reading the different scenarios we selected and organized  ICT-specific variables to characterize the ICT infrastructure and usage types. We distinguished three types of ICT-specific variables, the first being general ICT characteristics, with variables about the evolution of usages, about equipment (lifetimes, quantity of equipment, data center locations, use of cloud,...), and about usage (energy efficiency, data flows and data center consumption). All these characteristics directly impact equipment and infrastructures, and we call them \emph{infrastructure} variables. The second type, called \emph{applications}, addresses the kind of applications of ICT considered in the scenarios (teleworking, smart mobility, robots for manufacturing in industry,...), and the third type, \emph{technologies},  the digital technologies present in the scenario (data analytics, AI, cloud and edge computing,...).

Even if digital technologies exist in all scenarios, the studies rarely mention ICT-specific variables as driving the scenarios pathway. We observed that they are only discussed explicitly in quantitative studies that calculate energy consumption and greenhouse gas emissions (e.g., \cite{ademe2021transition,negaWatt,RTE}) and in prospective studies specifically targetting ICT as main topic (e.g., \cite{creutzig2022,CNIL,deron2022}).  In fact, as previously observed by \cite{creutzig2022}, no current conceptualization of decarbonization pathways explicitly accounts for the impacts of digitalization in the Anthropocene. 

Figure~\ref{fig:all_vars} gives the complete list of variables. The encoding of all scenarios according to our typology and variables is provided in the supplementary material~\footnote{\url{https://hal.science/hal-04362220v2/file/Supplementary_Analysis\%20of\%20ICT\%20in\%20prospective\%20scenarios.pdf }}.

\section{Results for ICT variables}
\label{sec:IT}
\begin{figure*}[h!]
    \centering  
     
    \begin{tabular}{cc}

    \includegraphics[width=0.49\textwidth,trim={0 0 0 0.9cm},clip]{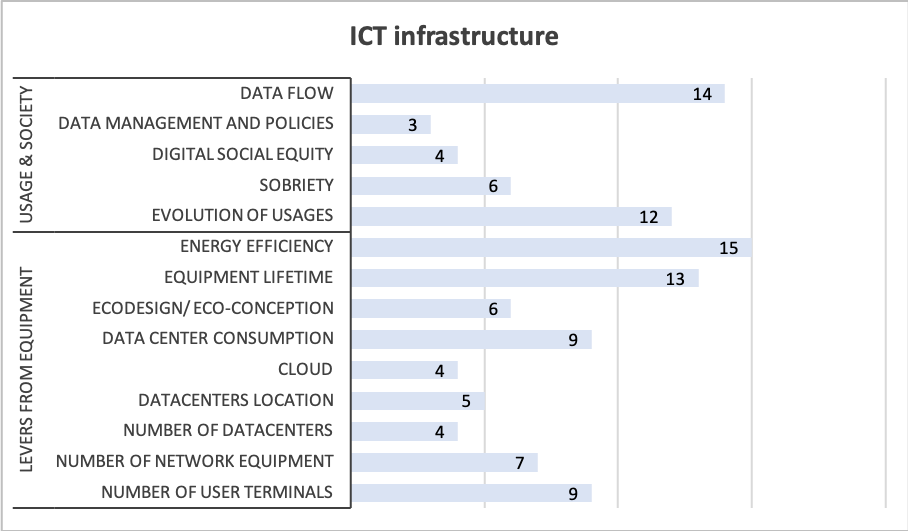}&\includegraphics[width=0.45\textwidth,trim={0 0 0 0.9cm},clip]{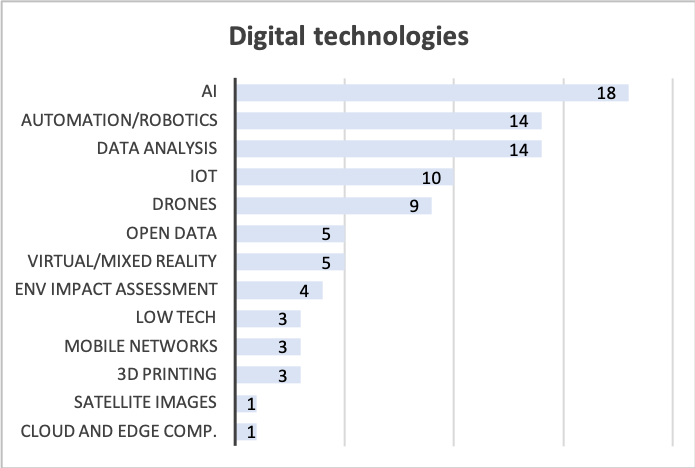}\\
    a) Infrastructures by scenario & b) Digital technologies by scenario 
    
    \end{tabular}
    \includegraphics[width=0.45\textwidth,trim={0 0 0 0.9cm},clip]{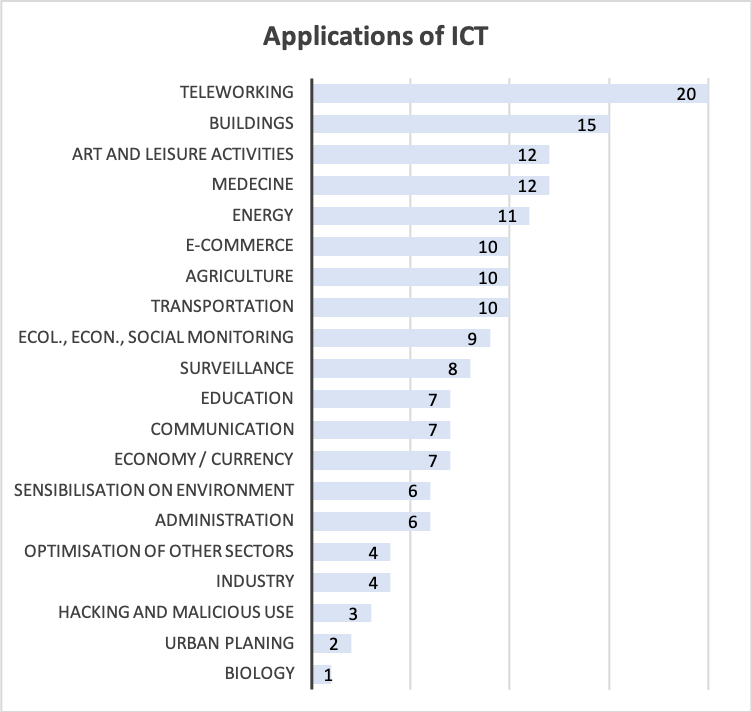}\\
          c) Applications by scenario \\
    \caption{Summary of our scenario analysis for ICT. Each diagram indicates the number of scenarios among the 35 explicitly mentioning an ICT variable. 
    For complete details, we refer the readers to our complementary material. }
    \label{fig:result}
\end{figure*}

We now present which digital technologies (section \ref{sec:techno}) can be found in prospective studies (section \ref{sec:appli}). All studies consider that ICT will be present in 2050 at different scales. In some scenarios, digital technologies are central in every aspect of society, but not in all of them. Here, we only present more frequent or surprising findings and refer the reader to supplementary material for complete details. Figure~\ref{fig:result} presents a summary of our findings.

\subsection{Digital technologies and ICT characteristics}\label{sec:techno}
Digital technologies that are the most present in scenarios are artificial intelligence (AI), robotics and automation, the Internet of Things (IoT), drones, and data analytics. Most studies explicitly mention AI and robotics for at least one scenario.  On the opposite, virtual and mix reality for working remotely and leisure can only be read in the stories of some narrative scenarios. Remark that technologies are not always directly named; we can only assume that this technology is used in a given application. For example, precise agriculture might be mentioned as an application but with no mention of AI and robotics within the scenario. In the complete table given in the supplementary material, we have tried to avoid making our assumptions. A technology is marked as present in a scenario only if explicitly mentioned.

As a consequence of the use of data by digital technologies, 5 scenarios mention open data in four studies: (in studies \cite{ademe2021transition,Arup, EUgreendeal, creutzig2022}). Open data is essential in the "Deliberate for the good" 
scenario of the ICT-specific study \cite{creutzig2022}. It comes with the idea that sustainable ICT requires more equity and collective data management.

Concerning the environmental impacts of digital technologies, four scenarios specifically mention mandatory life-cycle assessments of all new services and equipment (in studies \cite{Arup,Shift, EUgreendeal, deron2022}).

\subsection{Applications of digital technologies}\label{sec:appli}
As already said, digital technologies are present in every scenario we have read. Here, we present the applications for which these technologies are deployed. These applications are driven by general variables such as economic growth, demography, or lifestyles.

Note that scenarios are not entirely exhaustive regarding applications. It is not because an application is not explicitly mentioned that it does not  exist in the future that is described. Similar to what we did for technologies, we avoided making assumptions about omissions in the paper and supplementary material. 

\subsubsection{Main applications}

Almost all studies consider remote working, precise medicine and smart transports to be natural ICT applications. To a lesser extent, applications in farming, economy, and buildings are widely present. These applications are a natural extension of today's current development and funding of ICT. 

ICT is also considered in several scenarios for ecological, economic, and social monitoring. In particular, it enables tracking the state of the planet for forecasting and information on extreme events.

\subsubsection{Applications only discussed in a few scenarios}Some applications that already exist today are not discussed in all scenarios:
\begin{itemize}
    \item e-commerce: We only saw e-commerce in four studies (\cite{Shift,ademe2021transition,DDC}) with an increase (in particular for business-as-usual scenarios) or a decrease (for more sober scenarios). Supply chain management is mentioned in \cite{IPCC} but not specifically for e-commerce. 
    
    \item Information on climate change: ICT is important in some scenarios to raise awareness of climate issues and guide citizens. Digital media can be used to promote a culture of sustainability  \cite{SNBC,IPCC, EUgreendeal} and to inform about one's own emissions and carbon quota \cite{Eionet,Arup}.
    \item Social media: ICT is also still used for communication and social media and explicitly mentioned as such in the Regional Cooperations scenario of \cite{ademe2021transition}, in the Ecotopia and Technocracy scenarios of \cite{Eionet}, and in \cite{CNIL}. In these scenarios, digital technologies enable citizens to interact continuously.
    
    \item Leisure: Leisure relying on digital technologies is discussed in several studies, with different visions in different scenarios. In \cite{ademe2021transition}, the frugal scenario assumes less digital leisure, while the Restauration Gamble one assumes more and more digital leisure. Conversely, the \cite{negaWatt} scenario, which is a relatively sober scenario, supposes that digitalization continues for leisure activities. Even more surprising, the Ecotopia scenario of the study~\cite{Eionet}, which is, in essence, closer to nature, considers an increased presence of the digital sphere. 
    
    Digital technologies can also be used to improve the living environment, as in the Humans Inc. scenario of \cite{Arup}, which mentions digital walls displaying images of a clean waterfront to protect residents from unpleasant views.
    
    \item Surveillance: In general studies, surveillance was only found in the Extinction express scenario of \cite{Arup} with the use of micro-drones by the police and in centralized scenarios from \cite{DDC}, with AI or autonomous units. Naturally, it is discussed more deeply in~\cite{CNIL}, the main topic being data protection and individual privacy. As such, data surveillance is a central variable of this study.
\end{itemize}

Furthermore, while digital technologies are widely used in scientific domains (e.g., biology or physics) today, they are only mentioned in one study.

\subsubsection{Sobriety in digital technologies}

Some scenarios call for sobriety and therefore, bring sobriety within the digital sector inherently. This is visible, for instance, in the sufficiency scenario of \cite{RTE} with less digital leisure and less digital advertising. In the Frugal Generation scenario from \cite{ademe2021transition}, digital sobriety implies no new data center development and a stop of data growth. Another aspect of this sobriety concerns equipment repairability by users, which is discussed in many scenarios to increase equipment lifetime. In \cite{deron2022}, a reduction of the number of pieces of equipment is also possible with the development of equipment sharing in companies.

Some sober scenarios \cite{negaWatt, Shift} consider digital sobriety from a digital technologies usage angle. The proposal is to control usage and limit data flows. In the scenario from \cite{Shift}, digital sobriety includes reducing video resolutions or cloud gaming, or bounding equipment rate per person. But, on the contrary, this same scenario assumes an increase in the use of ICT in many sectors (such as higher education or defense). \cite{creutzig2022} propose an extensive application of ICT to match sustainable development goals. In \cite{deron2022}, business models around ICT are questioned to achieve more sobriety. A milestone specifically proposes to limit the influence of technology giants and to regulate the attention economy model.

\section{Discussion}
\label{sec:discussion}
In the previous sections, we analyzed the kind of ICT included in the scenarios according to the typology of variables we defined. This section gives more general feedback on the scenarios, particularly on what we considered missing concerning the place of ICT.

A first general finding is that every scenario involves ICT, but without discussing the challenges of using ICT in an environmentally-constrained world. However ICT has a growing environmental footprint~\cite{freitag2021}, and raises resilience issues that should be taken into account. This situation calls for the role of digitalization in the context of anthropocene to be reexamined, as \cite{creutzig2022} advocates.

In the following discussion, we examine how the studies considered these challenges and what is currently missing concerning ICT in prospective studies.

\paragraph{Lack of systemic view of ICT}

All scenarios admit a digital world quite similar to that of today. Although the imagined worlds can be more minimalistic or developed, sometimes with sobriety and efficiency, no significant change exists in their equipment and applications of IT. Some structural changes come from innovations in ICT, such as teleworking or autonomous driving.

As our goal was to study the place of digital technologies in future scenarios, we defined general ICT-specific variables to analyze if some scenarios consider them as the starting point. These variables encompass the growth of data, location of data centers, or sobriety usage. ICT variables (section \ref{sec:typology}) can be seen as parameters that can be changed to achieve different scenarios. They may serve as central hypotheses to design general scenarios, i.e., being one of the causes in the scenario. Two possible examples include 1) because there will be much efficiency in digital technologies in the future, these technologies will be more deployed; and 2) because of a limitation on the number of data centers, digital technologies have stopped being applied to some sectors of society. However, no general study has built upon these considerations. ICT variables remain only consequences of other aspects of these scenarios. While ICT-centered studies \cite{creutzig2022, CNIL, deron2022} build upon these ICT variables, their scenarios hardly mention the application of ICT in various areas of society.  Authors in~\cite{Eriksson2018} advise to reconsider how ICT affect not only direct users, but other stakeholders, the biosphere, and coming generations. This is lacking in most scenarios we have studied, but several scenarios propose to regulate usages, data economy, attention economy (e.g.~\cite{creutzig2022, Shift, deron2022}).

Scenarios rarely look at ICT with a comprehensive perspective, instead providing more often a citizen's point of view on ICT more than companies, organizations, or society one. Not having a more systemic view of society implies that ICT is very application-oriented, omitting all the interconnection with infrastructures and technologies. For example, \cite{Shift} offers a specific plan for the future of digital technologies, but separated from other sectors of activities and with few interactions. While digital sobriety is encouraged, ICT's place in other sectors is hardly discussed, even when it is already deployed.

In \cite{Erdmann2010}, the authors argued that 
forthcoming macroeconomic studies on ICT should account for  first-, second-, and third-order effects, reflect error margins resulting from data uncertainty and use scenario techniques to explore future uncertainty. However, the studies we analyzed generally do not, even when ICT-centered. Some studies point to possible rebound effects of digital technologies, but indirect effects of ICT and their assessment are never included in the scenarios.

\paragraph{Resilience and missing applications}
Many sectors that cover essential needs such as drinking, eating, housing, or health care depend on ICT today, and imagining their future is thus important to make the right decisions. No study covers them all concurrently yet, and for example supply chains or water supply management are never detailed. This lack of details raises questions regarding the transformations that an increase or decrease in digital technologies may imply for these applications and people working in these fields. 

In addition, ICT can be impacted by significant physical, resilience, and environmental challenges: ICT infrastructures, for example, are very vulnerable to extreme climate events. 
This vulnerability is very rarely evoked in the scenarios, and only \cite{Shift} has stated that "The ubiquity of digital technology makes its resilience crucial to many aspects of society". 
As digital technologies are often seen as crucial for decarbonization of other sectors, not considering their vulnerability may lead to inconsistent scenarios. 
The absence of a resilience variable thus constitutes a limitation of current studies.

Cybersecurity is another aspect that is rarely tackled in scenarios. The capacity of technologies to resist cyber-attacks is essential, especially in scenarios with more fragmentation of the world.

\paragraph{Questions related to ICT materiality}

Developing ICT in many sectors of activities, as proposed in some scenarios, would likely necessitate the installation of a significant amount of new digital infrastructures. Only some scenarios mention the availability of critical resources, pure water, energy, or land (e.g., with resource colonies on the Moon and deep sea mining in Extinction express of \cite{Arup}, or in \cite{RTE} scenarios for energy production). This physical reality behind digital technologies is most often ignored by the scenarios.

Relying on ICT in the future requires looking at energy and resource supply, which may be hindered by geopolitical tensions, collaborations between countries, and changes in governance. These tensions are discussed in French national studies (e.g. RTE and negaWatt) and in the narrative study Arup, but do not seem to influence how scenarios integrate ICT. The Extinction express scenario of the Arup study describes resource colonies on the Moon and in the deep sea. 

To the best of our knowledge, however, no study mentions the geopolitical aspects related to the location of data centers and the installation of ICT infrastructures (e.g., underwater cables or satellites).

\paragraph{New technologies}
Surprisingly, many disruptive technologies that have recently received attention and funding are not mentioned in any of the studies, which only build upon existing technologies. For example, few studies discuss 6G, and no studies discuss quantum informatics, intermittent computing, or DNA data storage. It is thus difficult to imagine how these innovations in ICT could be integrated into the scenarios. 

An example of a study where the disruptive technologies of cyborg botany and olfactory notifications are considered in scenarios was recently presented in 2022 by French designers\footnote{ \url{https://b-com.com/actualite/immersion-dans-le-numerique-responsable-de-2040}}. However, this work was not included in our analysis because of a lack of detail in the methodology. 

Significant changes in the use of technology, such as using low-tech solutions or studying collapse informatics~\cite{Tomlinson2013}, are also mostly absent from scenarios.

\paragraph{Disruption}
Similar to the challenges in new technologies, very few general scenarios are peripheral, i.e., include unlikely and extreme events and consider a discontinuous path to the future. Typically, these disruptive scenarios are narratives. 

In general scenarios, future ICT is quite similar to that of today. 
Our relationship to digital technologies is not deeply questioned, which may be why ICT is present in all scenarios without breakthrough innovations. 

It is surprising in the context of Anthropocene that none of the studies analysed here propose disruptive scenarios where ICT would be less present or central in the society. 
The LIMITS community has nevertheless been active to challenge business-as-usual scenarios in the context of scarcity for some years now. The reasons of this non-disruption of ICT in the scenarios would need to be more deeply explored.

\paragraph{Other environmental aspects}
The studies were all made from a climate change perspective, omitting biodiversity loss and other planetary boundaries, and focusing only on human futures without considering other species. These environmental problems may have other consequences on future scenarios. One that is directly related to ICT is water scarcity. Indeed, the manufacturing of ICT equipment and data center cooling have a substantial ecological footprint in terms of water use. Prospective studies on ICT should, therefore, broaden their narrations.

\paragraph{Global North centered}
Most innovations in the ICT sector and major industries of the domain come from Global North  countries. As a consequence, the studies we have chosen were made in OECD countries, which clearly has an influence on scenarios. Although some studies (e.g., Arup) 
have been designed by people from several countries worldwide, our work is most likely not representative of visions from Global South studies, and future work should consider how to imagine ICT for the future in Global South countries.

\section{Conclusion}
In this paper, we have analyzed the place of ICT in prospective scenarios. We have highlighted some variables that influence the development of digital technologies and the areas of applications where ICT might be essential in the future. Our analysis demonstrated that the feasibility and consequences of the increasing development of digital technologies are not sufficiently considered. Through our discussion, we have shown that the studies hardly question humanity's relationship to digital technologies, or the applications of ICT in human societies situated in a world of limits. The reason is probably that ICT is mostly seen as improving the society, as a solution to environmental challenges, or as immaterial.  This shortcoming limits the possibility of imagining more diverse plausible futures for the ICT sector.

There is therefore still a need to design prospective general and non-general studies that could offer more diverse and systemic views of ICT, and that could highlight which undone science~\cite{hess2015undone} should receive more attention. Prospective studies with a sustainability perspective for ICT can i) offer a more diverse and systemic view of the future of digital technologies, ii) highlight the need for discussing, structuring, or funding. As an example, \cite{deron2022} demonstrate that, based on prospective co-design workshops, it is possible to find milestones in order to reach a future that is desirable  of digital technologies. The milestones include among others sobriety measures, awareness, repairability and changes in some economic models, even if there is still work to be done to imagine the consequences on the others sectors of activities. They show samples of undone sciences~\cite{hess2015undone} for which social and industrial movements, funding, and research are needed for a desirable and sustainable ICT. 
Design fiction has been promoted for many years now by the LIMITS community (e.g.~\cite{Tanenbaum2016}) for its ethical, rhetorical, and playful role. It is a tool to play with utopias and dystopias, to open dialogue between scientific researchers and society, and to think about
the interplay between technological and sociological
futures. The design of further studies must also be accompanied with a better understanding of the gaps in public knowledge and the misconceptions about sustainability in order to be able to impact the narrative around collapse~\cite{Tanenbaum2016}. From our analysis, we additionally advocate for researching methodologies to propose more disruptive, and desirable for all, scenarios with a systemic point of view.

\bibliographystyle{ACM-Reference-Format}
\bibliography{biblio.bib}

\appendix

\clearpage
\section{Summary of studies}
\label{sec:studies}
    \topcaption{\normalsize Summary of studies reviewed for this paper}
    \label{tab:scenarios}
    {\footnotesize
\begin{supertabular}{C{1.2cm}p{6.2cm}}
         Study& Summary \& Scenarios \\
         \hline
        \multirow{2}{*}{\specialcell[t]{\emph{IPCC}\\2022}} & Report from IPCC Group III. This document does not correspond to a unique scenario, but describes the mitigation options.\\
        & \textit{IPCC}\\
        \hline
        
        \multirow{2}{*}{\specialcell[t]{\emph{Ademe}\\2022}}&  This study from French Agency for Ecological Transition ADEME proposes   five coherent and contrasting paths leading France towards carbon neutrality by 2050. \\ 
        & \textit{Business-As-Usual; Frugal Generation; Regional Cooperation; Green Technologies; Restoration Gamble}\\
        \hline
        
        \multirow{2}{*}{\specialcell[t]{\emph{negaWatt}\\2021}} & Scenarios for France made by the négaWatt association to achieve carbon neutrality by 2050 while reducing the extraction of raw materials. This scenario is also compatible with the -55\% greenhouse gas target set at European level for 2030.\\
        & \textit{negaWatt}\\
        \hline

        \multirow{2}{*}{\specialcell[t]{\emph{EU green deal}\\2019}} & This communication sets out the strategy to transform the EU into a fair and prosperous society, with a modern, resource-efficient and competitive economy where there are no net emissions of greenhouse gases in 2050 and where economic growth is decoupled from resource use. \\
        & \textit{EU green deal}\\
        \hline

          \multirow{2}{*}{ \specialcell[t]{\emph{Eionet} \\2022}}&This study produces a set of imaginaries offering engaging, plausible and contrasting images of what a sustainable Europe could look like in 2050. The imaginaries were developed through a participatory process, involving European Environment Agency staff, experts from the Eionet group on foresight, and external stakeholders. \\
          &\textit{Great decoupling; Ecotopia; Unity in adversity; Technocracy} \\
            \hline

        \multirow{2}{*}{ \specialcell[t]{\emph{Arup} \\2019}}& Study made by an international multi-disciplinary team of consultants, designers and scientists. In the study, Climate considerations come third, subordinate to economic development and societal wellbeing.  Scenarios are narratives and accompanied by a story told from the perspective of a person within that world, a timeline of events, and the achievement of the 17 UN SDGs. \\      
        & \textit{Greentocracy; Post Anthropocene; Extinction Express; Humans Inc.}\\
        \hline

        \multirow{2}{*}{ \specialcell[t]{\emph{DDC} \\2020}}& The Danish Design Center (DDC) designed the Living Futures: Scenario Kit for understanding, discussing, and shaping the future. The kit is built around a set of four alternative futures set in the year 2050  explored through narrated stories by fictive people living in them.  \\      
        & \textit{Centralised Market-driven; Centralised Society-driven; Distributed Market-driven; Distributed Society-driven}\\
        \hline
   .

        \multirow{2}{*}{ \specialcell[t]{\emph{SNBC}\\2020}}& The National Low Carbon Strategy (SNBC) describes a roadmap for France on how to steer its climate change mitigation policy. The SNBC is based on a reference scenario developed through a modelling exercise also used in the Multi-annual Energy Programming. The scenario details the public policy measures, in addition to those already in place, which will allow France to meet its short-, medium- and long-term climate and energy objectives as best it can. \\      
        & \textit{Baseline} \\
        \hline
        
        \multirow{2}{*}{ \specialcell[t]{\emph{RTE}\\2022 }} 
        &Study on the technical conditions necessary for a power system with a High Share of Renewables in France Towards 2050, under the RCP 4.5 trajectory from IPCC 5th report. \\ &\textit{Baseline; 
        Extensive Reindustrialisation;
         Sufficiency; 
        Acceleration 2030}\\
        \hline

        \multirow{2}{*}{ \specialcell[t]{\emph{Shift}\\2020 }} 
        &    The Shift Project published a plan for transforming French economy in 2022 for each sector of the economy. A preliminary report was published in end 2020 and this is the one we review here.  \\ &
        \textit{PTEF}\\
        \hline
     
        \multirow{2}{*}{\specialcell[t]{\emph{France 2072}\\2018 }}& This research work models two contrasting lifestyles for France in 2072,to answer the question: To what extent do lifestyles influence the energy system's capacity to achieve carbon neutrality?; And amongs other, is a digital world compatible with the need to decarbonize the energy system, as usually thought? \\
        & \textit{Digital society; Collective society}\\
        \hline
     
        \multirow{2}{*}{\specialcell[t]{\emph{D\&A}\\2022 }}& This Digitalization \& Anthropocene   research study presents the past, present, and future of digitalization. It presents three illustrative future pathways that span the possibility space for
digitalization and decarbonization in the Anthropocene. The authors conclude by identifying leverage points that shift human–digital–
Earth interactions toward sustainability\\
        & \textit{Planetary destabilization;Green but inhumane; Deliberate for the good}\\
        \hline
        
        \multirow{2}{*}{\specialcell[t]{\emph{CNIL}\\2021 }}& This study proposes new narratives and imaginaries regarding the protection of personal data in 2030 to help thinking regulatory practices. Throughout the process, highlights have been made on the use of technology in different social groups and at different times of digital life, as well as the resulting risks to individual and collective freedoms. \\
        & \textit{Renowed; Meddling; Home Sour Home}\\
        \hline
        
        \multirow{2}{*}{\specialcell[t]{\emph{Digit.}\\\emph{Challenge}\\2022 }}& The Université de Montréal and Espace pour la vie propose Chemins de transition ("Paths for transition"). This project combines a participative approach with major systemic challenges on a provincial scale. It combines different methodologies (backcasting, change-oriented approaches, transition management, etc.) and addresses 3 challenges: food, digital, territory. We here focus on digital challenge only. \\
        & \textit{Quebec 2040}\\
        \hline
\end{supertabular}}

\section{Scenario variables}

\begin{figure*}[h!]
\begin{center}
    \centering\makebox[0.95\textwidth]{
\begin{tikzpicture}[mindmap, grow cyclic, 
concept/.append style={fill={none}},
every node/.style={concept }, concept color=gray!60, 
level 1/.append style={level distance=3.5cm,sibling angle=110},	
level 2/.append style={level distance=3cm,sibling angle=45,  text width=2cm},	
level 3 /.append style={level distance=3cm,sibling angle=100,  text width=2cm}, 
every annotation/.style = {draw,fill = white}]
\node[scale=0.8, concept color=orange!80]{Variables}  [clockwise from=120]
child { node[scale=0.9, concept color=blue!80] {Goal and design}[clockwise from=180]
    child { node(Goal){Project goal}}
    child { node(Process){Process design}
    }
    child { node(Content){Scenario content}
    }
}
child { node[scale=0.9, concept color=blue!80]{Societal variables} [clockwise from=110] 
    child { node{Demography}}
    child { node(Lifestyles){Lifestyles}
    }
    child { node(Economy){Economy} }
    child { node(Governance){Governance}}
    child { node(NonIT){Non-digital tech.}
    }
}
child [level 2/.append style={sibling angle=75}] { node[scale=0.9, concept color=blue!80]{ICT variables} [counterclockwise from=195] 
    child[level 3/.append style={sibling angle=110,level distance=2.5cm}] { node(Infra){Infrastructures} [counterclockwise from=100] 
    child {node[scale=1.5](use){Usage \& Society}}
    child {node[scale=1.5](lev1){Levers from equipment}}
    }
    child { node(Appli){Applications}
    }
    child { node(Technologies){Technologies}
    }
}
;

\node [annotation, below left=-0.4cm and -1cm of Goal, scale=1.2, text width = 2cm] {\textbullet~ Inclusion of norms\\
\textbullet~ Vantage point\\
\textbullet~ Subject\\
\textbullet~ Time scale\\
\textbullet~ Spatial scales};
\node [annotation,above left=-0.4 and -1cm of Process, scale=1.2, text width = 3cm]  {
\textbullet~ Nature of the data\\
\textbullet~ Method of data collection\\
\textbullet~ Nature of the resources\\
\textbullet~ Nature of institutional conditions};

\node [annotation, right, scale=1.2, text width = 2.5cm] at  (Content.north east) {
\textbullet~ Temporal nature\\
\textbullet~ Nature of the variables\\
\textbullet~ Nature of the dynamics\\
\textbullet~ Level of deviation\\
\textbullet~ Level of integration};

\node [annotation,   below right =-0.5cm and -0.5 cm of Governance, scale=1.2, text width = 2.5cm] {
\textbullet~Importance of the state\\
\textbullet~Fragmentation of the world\\
\textbullet~Inequalities};

\node [annotation,  above right=-0.6cm and -0.6cm of Economy, scale=1.2, text width = 2.2cm]  {
\textbullet~Economic growth\\
        \textbullet~Decoupling\\
        \textbullet~Financing transition\\
        \textbullet~Re-industrialization};
        
\node [annotation, above right=-0.6cm and -0.6cm of Lifestyles, scale=1.2, text width = 2.2cm] {
\textbullet~Sobriety/frugality\\
\textbullet~Urban development\\
\textbullet~Work\\
\textbullet~Education\\
\textbullet~Type of farming};

\node [annotation, below right =-0.5cm and -0.5 cm of NonIT, scale=1.2, text width = 3.5cm] {
\textbullet~Resource circularity and availability\\
\textbullet~Decarbonation};

\node [annotation,  below right =-0.5cm and 2cm of Appli, below, scale=1.2, text width = 5.2cm] {

  \begin{minipage}{0.46\textwidth}
  \textbullet~Teleworking\\
\textbullet~Transportation\\
\textbullet~Industry\\
\textbullet~Surveillance\\
\textbullet~(Precise) medecine\\
\textbullet~Agriculture\\
\textbullet~Biology\\
\textbullet~Administration\\
\textbullet~Economy / Currency\\
  \textbullet~Art and Leisure \\
\textbullet~Energy
  \end{minipage}
  \begin{minipage}{0.52\textwidth}
\textbullet~Buildings\\
\textbullet~Urban planing\\
\textbullet~Monitoring\\
\textbullet~Optimisation of other sectors\\
\textbullet~Sensibilisation on environment\\
\textbullet~Communication\\
\textbullet~Education\\
\textbullet~e-commerce\\
\textbullet~Hacking and malicious use\\
  \end{minipage}
  };

\node [annotation, below right =-0.5cm and -0.6 cm of Technologies, scale=1.2, text width = 4.5cm] {
  \begin{minipage}{0.46\textwidth}
\textbullet~Data analysis\\
\textbullet~AI\\
\textbullet~Cloud/edge comp.\\
\textbullet~Automation/robotics\\
\textbullet~IoT\\
\textbullet~Drones
  \end{minipage}
  \begin{minipage}{0.52\textwidth}
\textbullet~Satellite images\\
\textbullet~Virtual/Mixed Reality\\
\textbullet~3D printing\\
\textbullet~Open data\\
\textbullet~Environmental impact assessment (LCA)\\
\textbullet~Mobile networks\\
\textbullet~Low-tech
  \end{minipage}};

\node [annotation,above left =-0.3cm and -1.7 cm of use, scale=1.2, text width = 2.8cm]  {
\textbullet~Evolution of usages \\
\textbullet~Sobriety\\
\textbullet~Digital Social Equity\\
\textbullet~Digital behavior changes \\
\textbullet~Data management and policies\\
\textbullet~Data flow	
};

\node [annotation,below right =-0.2cm and -1.7 cm of lev1, scale=1.2, text width = 2.8cm]  {
\textbullet~Number of user terminals\\
\textbullet~Number of Network equipment\\
\textbullet~Number of datacenters\\
\textbullet~Datacenters location\\
\textbullet~Cloud\\
\textbullet~Data center consumption
\textbullet~Ecodesign of equipment\\
\textbullet~Equipment lifetime\\
\textbullet~Energy efficiency};
  
  
\end{tikzpicture}}

\end{center}
\caption{Mindmap of variables describing the scenarios}
    \label{fig:all_vars}
\end{figure*}

\end{document}